\documentclass[journal]{IEEEtran}
\IEEEoverridecommandlockouts

\usepackage[cmex10]{amsmath}

\usepackage{enumitem}
\usepackage{graphicx}
\usepackage{color}
\usepackage{amsmath}
\usepackage{mathtools}
\usepackage{multicol}
\usepackage{multirow}
\usepackage[english]{babel}
\usepackage{blindtext}
\usepackage{algorithm}
\usepackage{algorithmic}
\usepackage{balance}
\usepackage{amsfonts}
\usepackage{bm}
\usepackage{stfloats}
\usepackage{subfig}
\usepackage{amsthm}
\usepackage{amssymb}
\usepackage{setspace}
\usepackage[nosort]{cite}
\usepackage{CJK}
\usepackage{cite}
\usepackage{caption}
\usepackage{comment}
\usepackage{array,multirow}
\usepackage{graphicx}

\theoremstyle{plain}

\usepackage[table]{xcolor}

\begin{document}

\captionsetup[figure]{labelformat={default},labelsep=period,name={Fig.}}

\title{Advancing Multi-Connectivity in Satellite-Terrestrial Integrated Networks: Architectures, Challenges, and Applications}

\author{Xiangyu Li,
        and Bodong Shang
\thanks{Xiangyu Li is with Shanghai Jiao Tong University, China and Eastern Institute for Advanced Study, Eastern Institute of Technology, Ningbo, China.}
\thanks{Bodong Shang (corresponding author) is with Eastern Institute for Advanced Study, Eastern Institute of Technology, Ningbo, China.}
}

\maketitle

\begin{abstract}
Multi-connectivity (MC) in satellite-terrestrial integrated networks (STINs), included in the Third-Generation Partnership Project (3GPP) standards, is regarded as a promising technology for future networks, especially the non-terrestrial network (NTN).
The significant advantages of MC in improving coverage, communication, and sensing through satellite-terrestrial collaboration have sparked widespread interest.
This article introduces three fundamental deployment architectures of MC systems in STINs, including multi-satellite, single-satellite single-base-station, and multi-satellite multi-base-station configurations.
Considering the emerging but still evolving satellite networking, we explore system design challenges such as satellite networking schemes, such as cell-free and multi-tier satellite networks.
Subsequently, key technical challenges severely influencing the quality of mutual communications, including beamforming, channel estimation, and synchronization, are discussed.
Furthermore, typical applications such as coverage enhancement, traffic offloading, collaborative sensing, and low-altitude communication are demonstrated, followed by a case study comparing coverage performance in MC and single-connectivity (SC) configurations.
Several essential future research directions for MC in STINs are presented to facilitate further exploration.
\end{abstract}


\IEEEpeerreviewmaketitle

\section{Introduction}
Fueled by the increasing demand for high-quality seamless connectivity and the rapid proliferation of data-intensive applications such as video streaming, cloud services, and the Internet of Things (IoT), the evolution of network architectures continues to advance inexorably.
\textcolor{black}{
Conventional terrestrial networks (TNs) can often be restricted by insufficient infrastructure and, thus, scarce spectrum resources, particularly in remote and underserved areas.
User equipment (UE) consumes radio resources from the same base station (BS) and the same radio access technology, known as carrier aggregation (CA); however, UE is still limited by the scarcity of bandwidth resources allocated by a BS.
This gives rise to the emergence of multi-connectivity (MC), a solution that enables simultaneous connections of a device to multiple networks to enhance the robustness, performance, and resilience \cite{3gpp.38.821}. Initially, this approach was introduced in Third Generation Partnership Project (3GPP) Release 11 and 3GPP Release 12.
Recently, the next generation of wireless communication systems, the sixth generation (6G), has embarked on extending the capabilities and scenarios of the fifth generation (5G), such as network coverage, traffic offloading, integrated sensing and communication, low-altitude economy, etc., to global IoT. 
}
By allowing devices to connect to multiple network nodes concurrently, connections between transmitters and UEs can be ensured with increased coverage, higher data throughput, and improved reliability. 
To satisfy this growing demand in more expansive areas, it is essential to understand the developments in potential architectures, technical challenges, and potential real-world applications of MC from a system-level perspective.

\textcolor{black}{
In recent years, due to the growing number and wider distribution of UEs, 
current TNs may be insufficient to provide high coverage and data rates with enough reliability. 
While TNs can be more widely deployed in urban areas, economic and geographic limitations have also presented challenges in resolving such problems by deploying TNs in rural and remote areas.
On the other hand, satellite communication (SatCom) is envisioned to be a perfect supplement for TNs because of its extensive coverage and efficient multicast and broadcast capabilities. 
However, challenges such as weather sensitivity, the high cost of deployment and maintenance, and deep fading at higher frequencies make it difficult for SatCom systems to support global services independently. 
Considering their respective advantages and limitations, satellite-terrestrial integrated networks (STINs) have attracted the attention of standardization institutes, network operators, research fraternities, and related industries \cite{sun2022integrated}. 
}
By reaping the benefits of the wide coverage provided by satellites and the high data rates of both networks, STINs act as a prospective paradigm for future networks in providing global coverage and ubiquitous wireless access. 
Note that if certain satellites or BSs become overloaded due to a large number of UEs, their resources are strained, while other satellites or BSs may remain underutilized. This imbalance leads to inefficient use of available network resources.

One promising solution to tackle such a problem is to redistribute the load by incorporating MC in STINs to switch or add UEs to less congested satellites and/or BSs. 
\textcolor{black}{
Using MC in STINs can bring substantial benefits to communications, including flexible bandwidth utilization, seamless handover stability, and efficient resource coordination and optimization.
Firstly, while no additional bandwidth is provided, UEs experience greater flexibility in connecting to multiple serving entities in the TN and/or non-terrestrial network (NTN). By exploiting MC in STINs, the excess available resources could be utilized and allocated more efficiently.
Secondly, the interoperability of satellites and TNs can support seamless handover and reduce the probability of handover failure, which will improve the reliability of the connection.
Thirdly, the information interaction among satellites and BSs can be made more efficient to realize more dynamic and adaptive space-ground resource coordination and optimization.
}
Note that MC in terrestrial mobile networks was once studied in terms of protocols, data, and control in the transport layer and the network layer \cite{pupiales2021multi}.
The authors in \cite{majamaa2024toward} also presented the concept of MC in NTNs but did not consider the inherent TNs.
An overview of the activities and goals of an ongoing STIN project was provided in \cite{morosi2024terrestrial}, without leveraging the advantages offered by MC.

\textcolor{black}{
It is worth mentioning that none of the above works considered the design and organization of different network elements from the perspective of large-scale system deployment. In addition, potential applications and key research directions related to next-generation networks should be envisioned for future development, all of which are left as valuable research gaps to be filled in.
Specifically, although MC in STINs is expected to offer potential benefits, several questions remain to be answered.
\begin{itemize}
    \item Q1: What will be the basic architectures for deployment when the MC is integrated into current networks?
    \item Q2: What will be the challenges to the system design of MC in STINs from a system-level perspective?
    \item Q3: What will be the real-world applications that MC will have in STINs for future communication systems?
\end{itemize}
Therefore, in this article, we seek to address these gaps in existing research and answer the aforementioned questions to direct the development of MC in next-generation networks.
}

\begin{figure*}[tp]
\begin{center}
\setlength{\abovecaptionskip}{+0.2cm}
\setlength{\belowcaptionskip}{-0.0cm}
\centering
  \includegraphics[width=7.0in, height=2.8in]{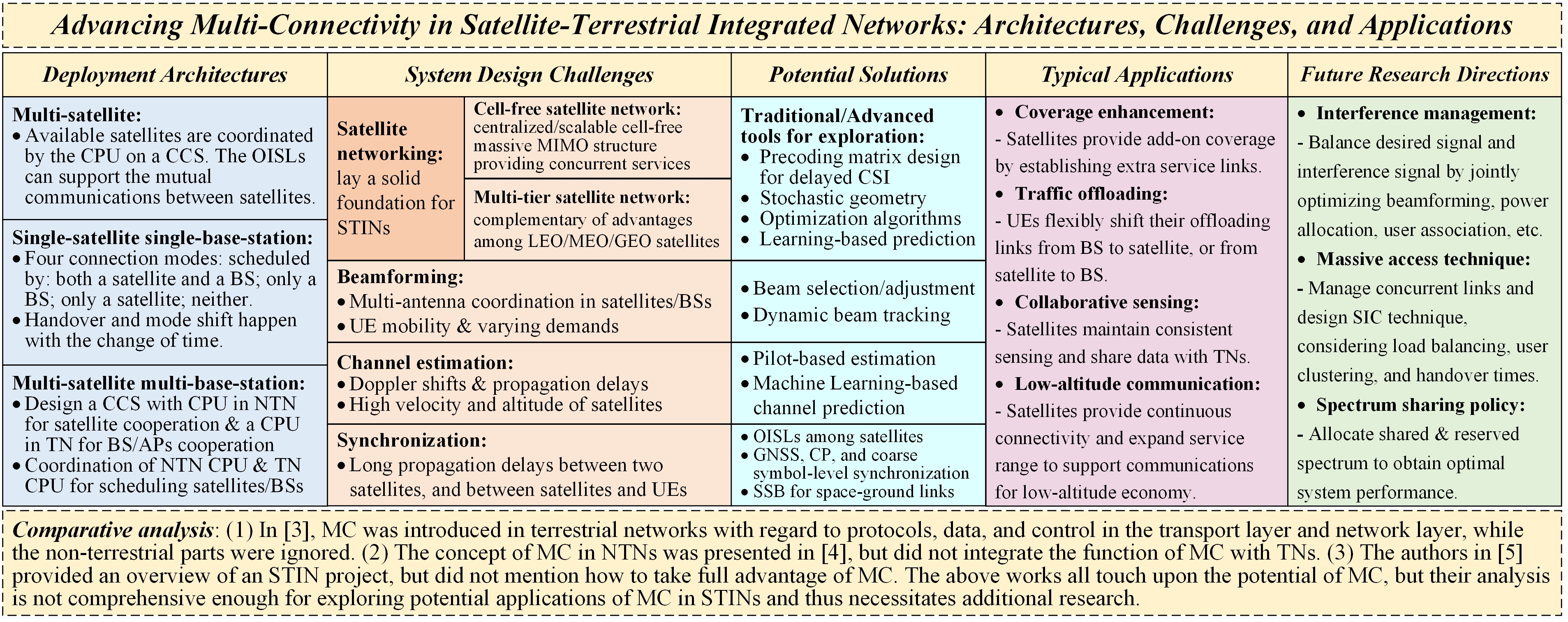}
\renewcommand\figurename{FIGURE}
\caption{\scriptsize \textcolor{black}{An overview of MC in STINs.}}
\label{fig:table}
\end{center}
\vspace{-8mm}
\end{figure*}

This article first provides an overview of MC in STINs in Fig. \ref{fig:table}. The main contributions are summarized as follows.

\captionsetup{font={scriptsize}}

\begin{itemize}
    \item We present three deployment architectures for MC when integrating satellite networks into current TNs. Importantly, we provide a comprehensive introduction to these architectures, detailing their characteristics and differences in user pairing and scheduling in each architecture.
    \item We investigate critical challenges in designing the MC system in STINs from a wireless networking perspective. Specifically, we elaborate on the challenges in the establishment of satellite networks, the use of transmission strategies, the application of channel estimation techniques, and the execution of synchronization operations.
    \item We propose four typical applications of MC in current and future networks, including coverage enhancement, traffic offloading, collaborative sensing, and low-altitude communication. Then, a case study is conducted to show the improvement in coverage when applying MC compared to single-connectivity networks.
\end{itemize}
Finally, the article concludes with a discussion on future research directions for exploring additional technical opportunities of MC in STINs.

\section{Deployment Architectures}
\begin{figure*}[tp]
\begin{center}
\setlength{\abovecaptionskip}{+0.2cm}
\setlength{\belowcaptionskip}{-0.0cm}
\centering
  \includegraphics[width=6.6in, height=4.4in]{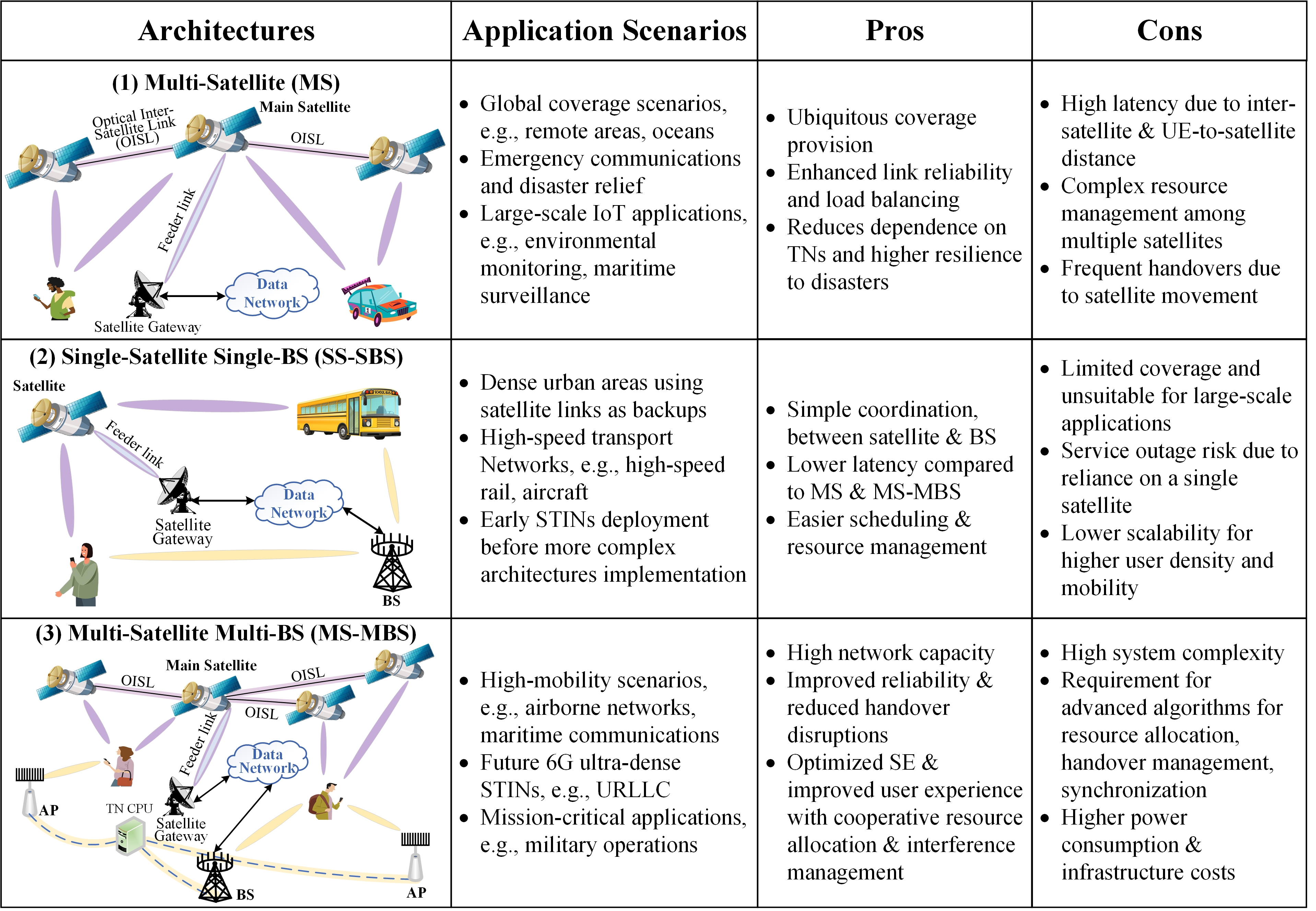}
\renewcommand\figurename{FIGURE}
\caption{\scriptsize \textcolor{black}{Deployment architectures of MC in STINs.}}
\label{fig:deployment_architecture}
\end{center}
\vspace{-8mm}
\end{figure*}

\textcolor{black}{
In the following subsections, we will discuss three typical deployment architectures of MC in STINs in terms of the working principle, application scenarios, and pros and cons, as shown in Fig. \ref{fig:deployment_architecture}.
}
\captionsetup{font={scriptsize}}

\subsection{Multi-Satellite (MS)}
Multiple satellite collaboration is a promising solution to enhance direct-to-cell communication.
In this architecture, the available satellites are coordinated by a central processing unit (CPU), which is deployed on one of the low-earth orbit (LEO)/medium-earth orbit (MEO)/geostationary-earth orbit (GEO) satellites with sufficient computing power and capabilities \cite{shang2023coverage}.
Specifically, the CPU can be equivalently referred to as a central server (CS) or a central control unit (CCU); the satellite with a CPU can also be referred to as the main satellite, the network controller (NC), or the central control satellite (CCS). The CCS is connected to the data network via the satellite gateway.
The optical inter-satellite links (OISLs) can be adopted for connections between satellites. The OISLs between the CCS and other satellites, more commonly known as backhaul links, convey the necessary control signals.

\subsubsection{\textcolor{black}{Working Principle}}
When a UE uses the MS architecture, it initiates a connection request to the CCS. 
The CPU on it then conducts necessary calculations and arranges a cluster of satellites to simultaneously transmit data to the UE in the same time-frequency resource block.
Clustering can be dynamic, and whether a handover is needed is determined by the received quality of service (QoS), the replacement of satellites in clusters, and the angle of the dome from the position of UE.

\textcolor{black}{
\subsubsection{Application Scenarios}
The MS architecture can be suitable for scenarios where terrestrial infrastructure is either unavailable or insufficient. It can provide global coverage in remote areas, oceans, and polar regions. In addition, it also plays an important part in emergency communications and disaster relief operations to ensure connectivity when TNs fail. Large-scale IoT applications, such as environmental monitoring and maritime surveillance, also benefit from this architecture.
}

\textcolor{black}{
\subsubsection{Pros and Cons}
With MS architecture, ubiquitous coverage can be ensured for areas beyond the reach of TNs. It also enhances link reliability and load balancing by selecting the best cluster of satellites. However, higher latency can be inevitable due to long propagation distances and inter-satellite coordination. It is also complicated to manage spectrum and power resources among multiple satellites. Moreover, satellite movement leads to frequent handovers and thus degrades service continuity. 
}

\subsection{Single-Satellite Single-Base-Station (SS-SBS)}
To further incorporate BS in TNs, the SS-SBS architecture can act as a foundation for network extensions to STINs.
Note that this scenario indicates the potential availability of both a satellite and a BS for connection. Based on the time-division strategy, there are three modes of connection: Mode 1, where the UE is scheduled to be served by both a satellite and a BS; Mode 2, where the UE is only scheduled to be served by a satellite; Mode 3, where the UE is only scheduled to be served by a BS; and Mode 4, where the UE is neither scheduled by a satellite nor a BS \cite{shang2024multi}.

\subsubsection{\textcolor{black}{Working Principle}}
A pair of available satellite and BS is initially selected to set the UE's connection into Mode 1. 
The availability of the current satellite is determined by the transmission blockage and its inner mechanism switchover.
In contrast, the availability of current BS is due to transmission blockage, unsatisfactory signal power, or UE's movement between cells. All of the above events will lead to handover.
In addition, with the change of time, this connection status may not persist due to mobility of the current satellite and UE, and may thus shift to Mode 2, 3 or even 4.

\vspace{-4mm}
\textcolor{black}{
\subsubsection{Application Scenarios}
The SS-SBS architecture can be applied to dense urban environments, where satellite connectivity serves as a backup to complement TNs. It can also be adopted for high-speed transportation networks, such as trains, aircraft, and ships, to enhance coverage along routes with limited terrestrial infrastructure. During the early stage of STINs, this architecture can be constructed before developing specific and more complex architectures.
}

\vspace{-4mm}
\textcolor{black}{
\subsubsection{Pros and Cons}
With only one satellite and one BS, the coordination between the two is comparatively simpler. Compared to the MS architecture, this architecture requires fewer handovers and thus enjoys lower latency. 
Additionally, scheduling and resource management are also straightforward. 
However, as coverage by one satellite is limited, this architecture may be unsuitable for large-scale applications.
The risk of service outages can also increase due to its reliance on a single satellite.
Furthermore, this architecture lacks scalability and cannot efficiently support a high density of UEs or high-mobility scenarios.
}

\subsection{Multi-Satellite Multi-Base-Station (MS-MBS)}
The MS-MBS architecture is an ideal but generalized combination and extension of the above two setups.
It not only covers the MS architecture but also extends the SS-SBS architecture.

\subsubsection{\textcolor{black}{Working Principle}}
To reduce complexity in NTN-TN integration, one of the satellites is designed as a CCS to coordinate with the terrestrial core network via the NTN gateway.
Unlike SS-SBS that coordinates the serving satellite and serving BS directly via the gateway, the MS-MBS architecture requires joint coordination of the NTN CPU and the TN CPU, as they each collect information of their cluster of satellites or ground BSs/APs.
\textcolor{black}{
Herein, the BSs include not only terrestrial gNB stations in large-scale networks, but also distributed access points (APs) in small-range networks.
All these BSs, i.e., gNB stations and APs, are within a certain area and are connected to a terrestrial TN CPU for coordinated transmission administration. 
In future networks, a TN CPU will enable the seamless integration of terrestrial gNB stations and APs to form a cohesive and unified network.
}

\textcolor{black}{
\subsubsection{Application Scenarios}
The design of the MS-MBS architecture can be expected to support high-mobility UE scenarios, where seamless and reliable connectivity is of importance. In typical scenarios including airborne networks and maritime communications, multiple satellites and BSs dynamically coordinate to guarantee uninterrupted service.
This architecture is also essential for future 6G ultra-dense STINs and can support ultra-reliable low-latency communications (URLLC).
Furthermore, mission-critical applications, such as military operations and space exploration, can leverage MS-MBS architectures for their redundancy, scalability, and fault-tolerant capabilities.
}

\textcolor{black}{
\subsubsection{Pros and Cons}
In general, allowing UEs to maintain multiple simultaneous connections can result in significant benefits, including near-maximized capacity, seamless connectivity, and improved network reliability. With cooperative resource allocation and interference management, SE can be optimized, and user experience can be improved. 
However, the main challenge lies in the complexity of coordinating multiple satellites and BSs, as advanced algorithms are required for synchronization, resource allocation, and handover management. In addition, compared to MS and SS-SBS, this architecture can have much higher infrastructure and operational costs.
}

\section{System Design Challenges}
In this section, we discuss four key design challenges from a system-level perspective, i.e., satellite networking, beamforming, channel estimation, and synchronization. 
\textcolor{black}{
Specifically, we first introduce two candidate networking schemes for satellites. Then, beamforming strategies are discussed to improve the quality of desired signals and at the same time suppress interference signals. Next, considering unknown and long propagation conditions, reasonable channel estimation methods need to be applied. Finally, an important factor, i.e., synchronization, is analyzed to guarantee the robustness of the overall MC system.
}

\subsection{Satellite Networking}
The iteration of each generation network needs to satisfy compatibility requirements.
While the TN architecture has been involved in the past few decades and cannot be easily changed, the satellite network and its integration with TN are still at an early stage.
Therefore, appropriate satellite networking schemes lay the foundation for STINs.

As mentioned, the available satellites should be appropriately organized to perform cooperative transmissions.
Depending on the cross-layer constellation designs, multiple satellites can be coordinated in various ways. Two promising schemes are listed below: cell-free (CF)-enabled networking and multi-tier-enabled networking.

\subsubsection{Cell-Free Satellite Network}
CF massive multiple-input multiple-output (MIMO), initially proposed in TNs, is anticipated to provide a new paradigm for satellite communication systems. In a traditional CF-enabled network, all APs concurrently serve all UEs in the same time-frequency resource block via time-division duplex (TDD) operation. 
Recently, scalable CF massive MIMO systems have been deemed more practical, with user-centric deployment through AP clustering emerging as a potential solution.
Each UE has its own set of APs for services; therefore, each AP collaborates with different APs when serving different UEs.
Similarly, organized in a CF and scalable manner, satellites can be dynamically clustered to collaboratively communicate with terrestrial UEs at different locations, especially those with poor propagation conditions \cite{zheng2024mobile}. In this case, the CF structure can be leveraged in future ultra-dense LEO satellite networks to provide a macro-diversity gain for higher spectral efficiency (SE), energy efficiency (EE), and network flexibility \cite{li2025downlink}. 

\subsubsection{Multi-Tier Satellite Network}
The multi-tier satellite networks, or multi-altitude satellite networks, briefly refer to satellites distributed evenly across inclined circular orbits at different altitude levels \cite{okati2023stochastic}. 
They take advantage of the distinct characteristics of satellites positioned at different altitudes of LEO, or MEO and GEO, and dynamically adjust their roles based on locations and specific service requirements of UEs. 
LEO satellites can provide high-speed data services and low-latency connections for UEs in urban areas, while MEO or GEO satellites support wider terrestrial coverage and navigation purposes. 
This mutual collaboration forms a coordinated network that offers complementary coverage and improved service delivery. 

For example, when the UE experiences poor propagation conditions with the current LEO satellites, the satellite network can accurately locate the UE and seamlessly switch its connections to LEO satellites at a different altitude or position to ensure uninterrupted service.
By employing this multi-altitude networking scheme, satellite networks can be more resilient to changes in UE density, interference, or environmental factors such as weather conditions. Moreover, as satellites can share the load based on their altitude-specific capabilities, more efficient resource management can be achieved, and the overall SE and EE of ultra-dense satellite networks can be improved.

\textcolor{black}{
By adopting the MC technique, a UE can be provided with more desired signal sources, improving not only its own performance but also the performance of the system.
}
However, implementing the above two satellite networking schemes can be challenging due to the high mobility of satellites. Although mutual information can be exchanged via OISLs between satellites, especially between the CCS and one of the other serving satellites, the channel state information (CSI) obtained during uplink training may not match the actual downlink transmissions.
\textcolor{black}{
One possible solution is to minimize the harmful effects of delayed CSI via a precoding technique that uses the long-term properties of CSI uncertainty \cite{omid2024tackling}. 
The satellite network can be preliminarily established and analyzed using stochastic geometry. 
Moreover, the MC technique can obtain more long-term property data for a better precoding matrix design. By incorporating learning-based channel prediction techniques and optimization algorithms, this solution may be more computationally efficient with higher precision.
}

\subsection{Beamforming}
Beamforming has long been an effective strategy for achieving the desired directionality and gain.
When designing MC systems in STINs, beamforming needs to be controlled so that the directions of multiple signals can be targeted at specific UEs or areas at almost the same time. This process can make full use of the desired signals received by UEs.
In addition, by controlling the directions of transmitted signals, interference signals received by UEs can be mitigated, especially in environments where UEs are densely distributed.
This will guarantee more reliable and high-quality communications and better user experience, and thus improve effective coverage and network throughput.

\captionsetup{font={scriptsize}}
\begin{figure}[tp]
\begin{center}
\setlength{\abovecaptionskip}{+0.2cm}
\setlength{\belowcaptionskip}{-0.0cm}
\centering
  \includegraphics[width=3.4in, height=2.9in]{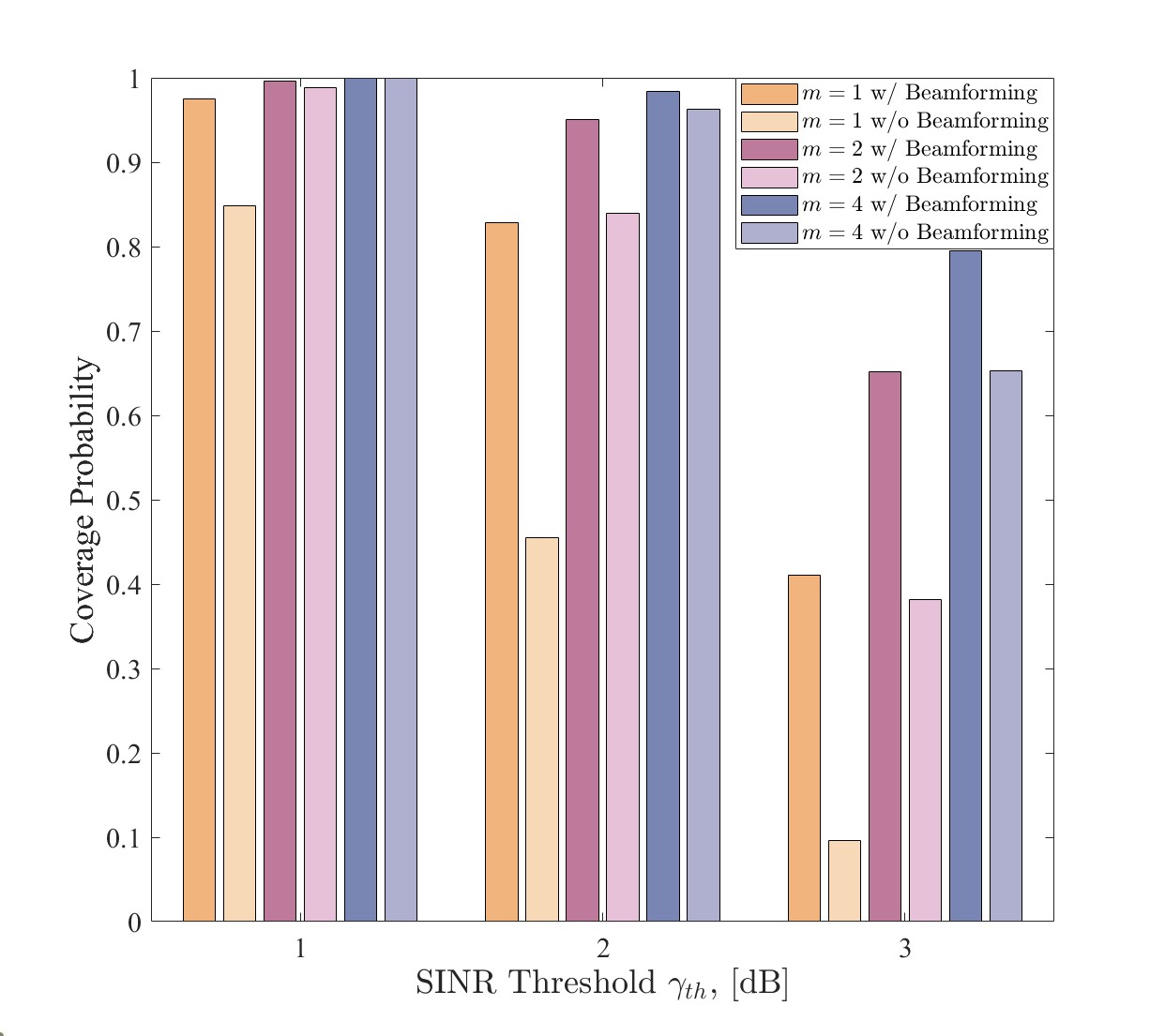}
\renewcommand\figurename{FIGURE}
\caption{\scriptsize Coverage probability versus SINR thresholds with different Nakagami-$m$ parameters. 
The satellites are distributed with the density $\lambda_S=1\times 10^{-5}$/km$^2$ at an altitude of $H_S=500$ km. The transmit power, antenna gain, and pathloss exponent are $\rho_d=50$ dBm, $G_t=30$ dBi, and $\alpha_N\approx 2.0$, respectively.
The UEs are distributed on the earth surface with the density $\lambda_U=4\times 10^{-6}$/km$^2$, and the receive antenna gain is $G_r=10$ dBi.
The cases of beamforming and non-beamforming are compared.}
\label{fig:BF}
\end{center}
\vspace{-6mm}
\end{figure}

\textcolor{black}{The benefits of beamforming can be shown in Fig. \ref{fig:BF}, where the effects of beamforming and non-beamforming configurations are compared for an MS architecture under different signal-to-interference-plus-noise ratio (SINR) thresholds, and $m$ is the parameter for \textcolor{black}{Nakagami-$m$ fading\footnote{\textcolor{black}{In satellite communications, the Nakagami-$m$ distribution offers versatility in modeling a wide range of small-scale fading phenomena for different satellite links. By changing its parameter $m$, it is possible to model the fading conditions of the signal from severe to moderate and fit the distribution to empirically measured fading data \cite{shang2023coverage}.}}}}. 
It is shown that beamforming counts for the improvement in coverage performance.
However, after integrating satellites and BSs, beamforming becomes more complex because multiple antennas need to be coordinated cooperatively. This problem intensifies when the beam patterns are dynamically adjusted in response to UE mobility and their varying demands.

\textcolor{black}{
Potential solutions can be further explored to allow MC to utilize multi-paths to improve signal quality. 
For example, beam selection and adjustment can be applied to optimize transmission paths by choosing beams with the best channel conditions to ensure stronger received signals.
Dynamic beam tracking can continuously adjust beam directions to follow UEs' movements, reducing link failure and improving reliability. 
Moreover, coordinated beamforming across satellites and BSs helps mitigate interference and enhance SE.
}

\subsection{Channel Estimation}
Channel estimation is also crucial for MC in STINs, as inaccurate CSI will downgrade the efficiency of uplink and downlink transmissions. 
In the uplink stage, the satellite/BS employs multi-user detectors to retrieve the data symbols sent by UEs to reduce multi-user interference; in the downlink stage, the satellite/BS employ techniques such as precoding, beamforming, or MIMO and each UE adopts estimated channels to demodulate and decode the received signal. 
These implementations heavily depend on the CSI acquired at both the satellite/BS and UE, respectively.
To harness the corresponding performance gains, satellites, BSs and UEs need to obtain CSI with sufficient accuracy \cite{li2023channel}.
Furthermore, MC requires UEs to establish multiple communication links, encompassing those in both the \textcolor{black}{NTN and TN}.
\captionsetup{font={scriptsize}}
\begin{figure}[tp]
\begin{center}
\setlength{\abovecaptionskip}{+0.2cm}
\setlength{\belowcaptionskip}{+0.2cm}
\centering
  \includegraphics[width=3.4in, height=2.7in]{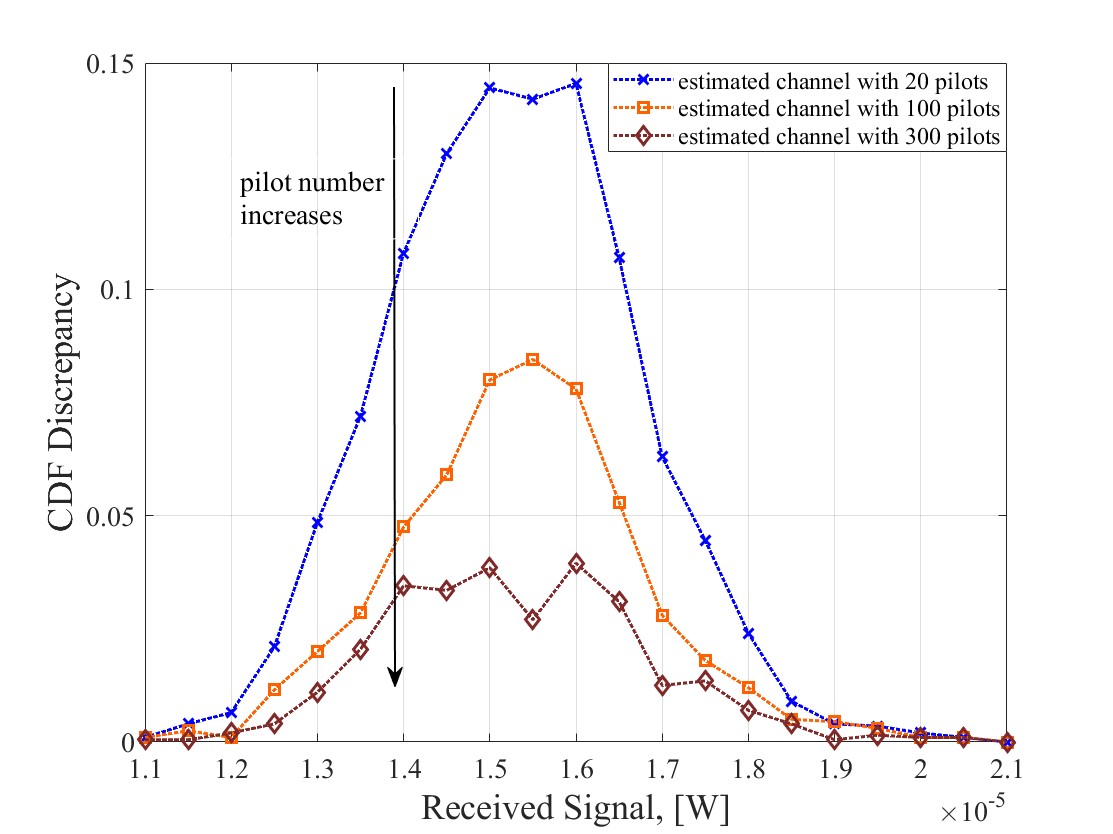}
\renewcommand\figurename{FIGURE}
\caption{\scriptsize Discrepancy between CDF of the signal received in the ideal channel and that in the estimated channels, with different numbers of pilots. The simulation parameters are the same as those in Fig. \ref{fig:BF}, except that $m=2$.}
\label{fig:Pilot}
\end{center}
\vspace{-6mm}
\end{figure}

\textcolor{black}{
Despite channel estimation challenges in STINs, MC can also help improve CSI accuracy through multi-source data, i.e., data from NTN and TN links. Specifically, by aggregating CSI from multiple BSs and satellites, MC provides a more comprehensive channel characterization, which can reduce estimation errors.
Correspondingly, with more data obtained, machine learning (ML)-based CSI prediction could be more precise by analyzing multi-source historical CSI patterns.
}

\textcolor{black}{
In addition, pilot-based estimation can be used in the STIN MC system.}
For example, during the downlink process, pilot signals, which are pre-known to UEs, are periodically inserted prior to the data symbols that are intended for transmission by satellites and/or BSs. 
Generally, if an adequate number of orthogonal pilots are available, ensuring that not too many UEs share the same pilot, pilot contamination can be mitigated, and the accuracy in channel estimation will be enhanced. 
\textcolor{black}{Simulation results in Fig. \ref{fig:Pilot} illustrate how the number of pilots influences the discrepancy between the cumulative distribution function (CDF) of the signal received in the ideal channel and that in the estimated channels.} 
The more accurately the channel is estimated, the smaller the CDF discrepancy will be. With an increasing number of pilots, the discrepancy narrows, which shows that the effects of pilot contamination can be gradually alleviated.

\subsection{Synchronization}
\textcolor{black}{
The last but not least important factor that makes cooperative transmission challenging for practical implementation is synchronization. Synchronization, which is also known as compensation for time delays, mainly results from the long-varying propagation delays dependent on the satellite-direct-to-UE distance.
}

\textcolor{black}{
To provide high QoS for UEs with sufficient reliability, it is crucial to maintain robust downlink synchronization between the UEs served, satellite(s) and/or BS(s) at the time of transmission. This can be achieved by designing the specific synchronization signal block (SSB) for each deployment architecture.
In the following, we will present and analyze synchronization solutions for the three previously discussed deployment architectures.
}

\subsubsection{\textcolor{black}{Synchronization for MS}}
Using UE positioning data from the Global Navigation Satellite System (GNSS) and the ephemeris data of each coordinate satellite, the CCS performs coarse frequency shift compensation for the data streams from each satellite. It also manages the delay differences of these streams to ensure that they remain within the length of the cyclic prefix (CP). This process enables coarse symbol-level synchronization between data streams \cite{yang2023distributed}.

\subsubsection{\textcolor{black}{Synchronization for SS-SBS}}
When a ground UE requests cooperative transmission from an NTN satellite and a TN BS, the timing advances are calculated based on known factors such as the UE location, the BS location, and satellite ephemeris. 
The satellite first calculates its timing advance to the UE according to the transmission time slot of its cooperative BS. Then, the satellite and BS select different transmission time slots so that signals from the satellite and BS arrive at the UE simultaneously \cite{shang2024multi}.

\subsubsection{\textcolor{black}{Synchronization for MS-MBS}}
Similarly but with a step further, the CCS in NTN and the CPU in TN need to work closely for the more complex MS-MBS architecture.
They first share UE positioning and CSI to establish a common consensus on the UE environment. Then, they collaboratively determine the best transmission strategy, including resource allocation and load balancing. Next, the CCS manages the coarse synchronization of the satellite signals, while the TN CPU aligns the timing of BS transmissions. Fine synchronization adjustments ensure that signals from satellites and BSs arrive at the UT with minimal delay discrepancies. By exchanging control information over dedicated links, the NTN CCS and the TN CPU coordinate the alignment of data streams to maximize signal quality at the UE. Finally, a UE feedback loop provides real-time CSI, allowing both entities to adjust transmission parameters dynamically. 
This collaboration process controlled by the NTN CCS and the TN CPU ensures that transmissions are effectively synchronized even with varying channel conditions and mobility of the UE and satellites.

In summary, for TN-NTN joint transmissions, the exchange of signaling for CSI and transmitted signals is required. A trade-off must be struck between the performance gains of the system and the complexity of implementation.

\section{Typical Applications of MC in STINs}
\begin{figure*}[tp]
\begin{center}
\setlength{\abovecaptionskip}{+0.2cm}
\setlength{\belowcaptionskip}{+0.2cm}
\centering
  \includegraphics[width=6.5in, height=4.8in]{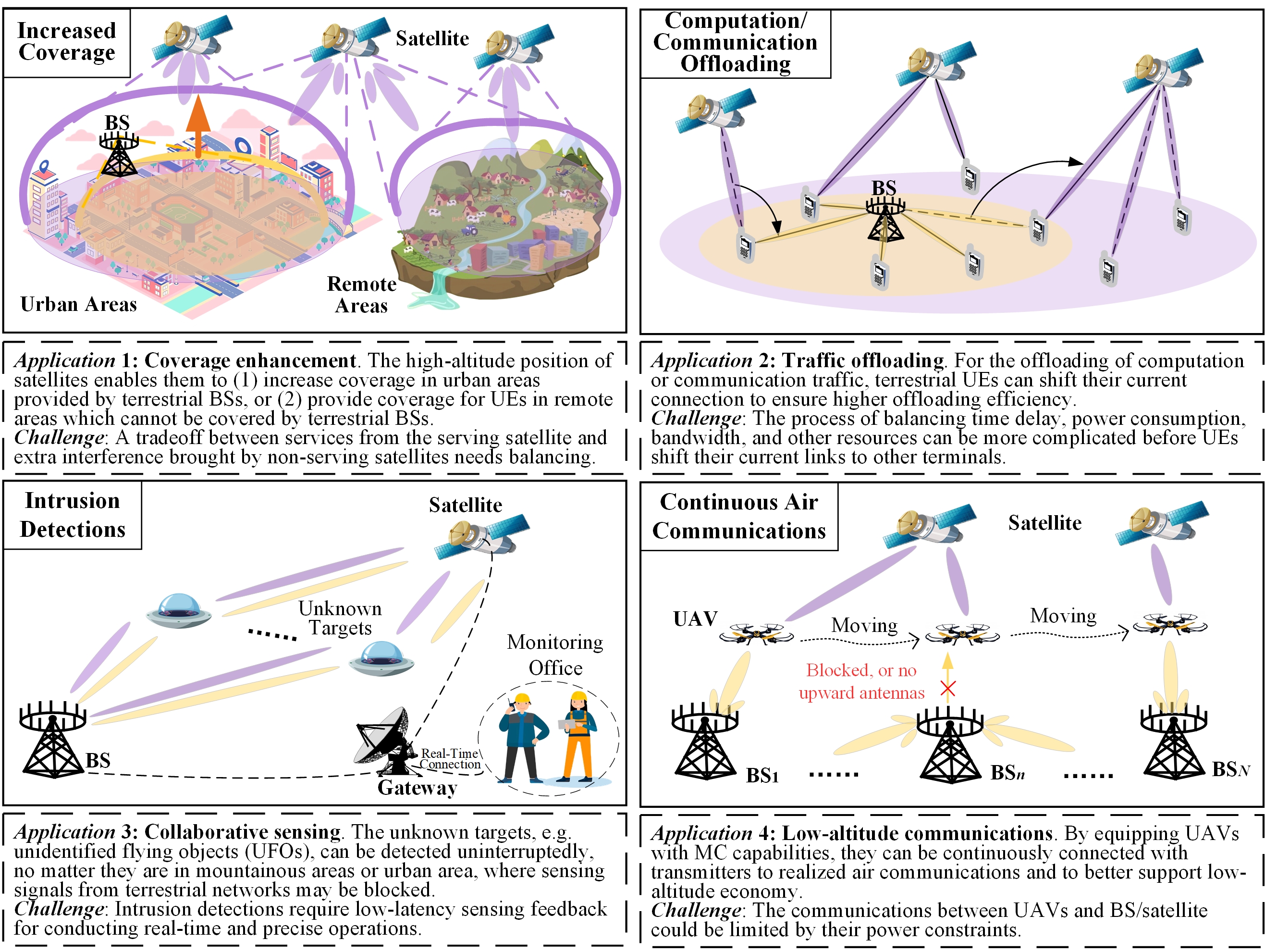}
\renewcommand\figurename{FIGURE}
\caption{\scriptsize A sketch of application scenarios of MC in STINs.}
\label{fig:applications}
\end{center}
\vspace{-8mm}
\end{figure*}

In this section, we present four typical applications that enable MC in STINs, as shown in Fig. \ref{fig:applications}, including coverage enhancement, traffic offloading, collaborative sensing, and low-altitude communication.
\captionsetup{font={scriptsize}}

\subsection{Coverage Enhancement}
Due to the high cost and complexity of infrastructure deployment and maintenance, TNs, especially where traditional infrastructure faces geographical limitations, can often struggle to provide ubiquitous and sufficient coverage in remote areas with few BS or urban areas with heavy communication requirements. 
However, satellites can bridge this gap by providing complementary coverage to ensure that UEs in areas of high traffic or hard-to-reach areas remain connected. 

Specifically, in urban areas, compared with single satellite communications, MC takes advantage of the existing TNs and reaps the ``add-on'' coverage by establishing communication links between satellite(s) and the UE who have already been served.
In remote areas where terrestrial BSs are sparsely distributed or even hardly available, the UE could resort to multiple satellites for service provision instead of relying only on a single satellite \cite{bai2018Multi}. 
Combining satellites and terrestrial BSs with MC technology will guarantee more robust and expansive coverage for UEs served.

\subsection{Traffic Offloading}
\textcolor{black}{
The increasing number of UEs and the booming number of communication and computation-intensive services have led to increased traffic restrictions. 
It is necessary to widely adopt traffic offloading because with the integration of networks, communication and computation-intensive services can overwhelm UEs, leading to increased latency, higher energy consumption, and overall performance degradation.
On the other hand, while UEs can handle some tasks locally, offloading allows the heavy processing to be shifted to more powerful edge or cloud servers, reducing the burden on UEs themselves.}

In MC-enabled STINs, UE is designed to support satellite-terrestrial dual-mode communication, so that connection shifts can be realized.
\textcolor{black}{
For practical implementation cases, for instance, in urban areas with densely distributed UEs, edge computing servers can allocate communication and/or computation tasks dynamically between satellites and BSs to guarantee load balancing and service continuity. 
These UEs may also offload their tasks, partially or completely, to available satellites, according to their channel conditions, traffic amount, etc.
Meanwhile, in remote or rural areas where TN coverage is limited, MC can shift the UE's connection from a BS to an available satellite to maintain connectivity.
}

\textcolor{black}{
However, the implementation of traffic offloading can be faced with technical challenges, including real-time link quality estimation, handover latency, and inter-network synchronization. To improve offloading efficiency, artificial intelligence (AI)-driven scheduling algorithms will be employed to predict traffic loads and reassign connections.
}

\subsection{Collaborative Sensing}
MC also plays an important role in collaborative sensing, especially for the detection of unauthorized entries and unknown targets.
For instance, when unknown targets are detected, MC can provide seamless sensing capabilities in diverse environments such as mountainous and urban areas. On the contrary, BSs that perform sensing tasks in TNs can often be confronted with signal blockages due to irregular natural environments and dense buildings.
By encompassing satellite links to provide additional detection feedback, robust and continuous sensing tasks can be guaranteed \cite{li2024bistatic}.

To be specific, while terrestrial BSs may lose line-of-sight (LoS) and experience signal degradation when the target moves within areas with densely forested mountainous or urban areas, satellites will maintain consistent sensing data streams due to their wide coverage. Sensing results can be shared with BSs via the satellite gateway and the data network. 
Although satellite sensing alone is less accurate than integrated sensing empowered by MC, general locations of the targets can be continuously recorded. When a BS regains its echo signals, MC-enabled sensing will be reactivated. 
This will ensure real-time tracking of targets regardless of their locations, and further enhance the reliability and precision of related intrusion detection systems.

\subsection{Low-Altitude Communication}
Due to the rise of low-altitude economy in recent years, effective air-ground coordination requires reliable and real-time communications with low-altitude platforms such as unmanned aerial vehicles (UAVs), weather balloons, etc.
In such cases, continuous connectivity to transmit data such as sensor readings, video information, and operating records to terrestrial stations is a vital prerequisite. 
However, when ground receivers lack upward-facing antennas or when upward-facing antennas are blocked in some directions, data transmission could be incomplete. MC can ensure uninterrupted communication by seamlessly switching UEs to available ground BSs and/or satellites.
This can be extremely valuable for low-altitude communication in rural, remote, or maritime areas with few terrestrial infrastructures.

In addition, MC can also facilitate the expansion of the service range of low-altitude communications. As the number of low-altitude platforms grows, traffic management in the airspace remains a critical problem. 
MC can be used to improve both operational efficiency and regulatory compliance by navigating multiple moving platforms in a more organized and real-time manner without discontinuity of echo signals.
This will facilitate the expansion of low-altitude economic activities, enable businesses to explore new service areas, and improve the quality and reliability of their offerings.

\section{Case Study: Multi-Connectivity versus Single-Connectivity for Coverage Enhancement}

\captionsetup{font={scriptsize}}
\begin{figure}[tp]
\begin{center}
\setlength{\abovecaptionskip}{+0.2cm}
\setlength{\belowcaptionskip}{-0.0cm}
\centering
  \includegraphics[width=3.4in, height=2.7in]{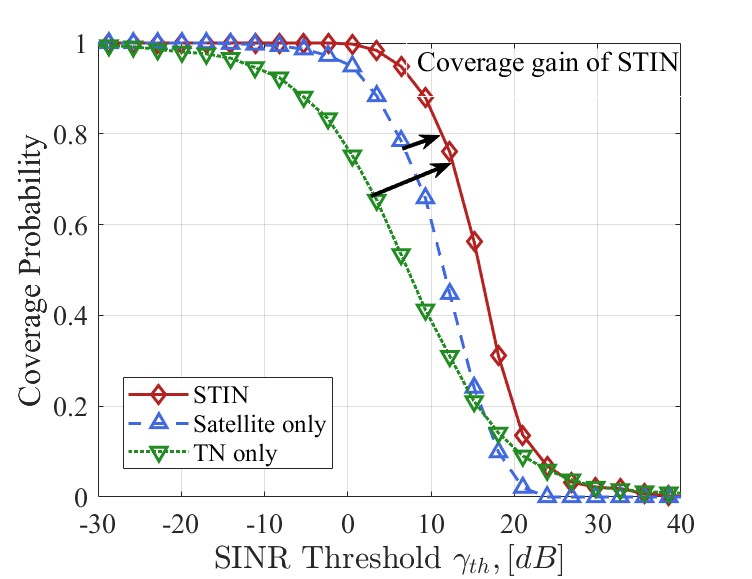}
\renewcommand\figurename{FIGURE}
\caption{\scriptsize Coverage probability against varying SINR thresholds under different connectivity schemes. 
The coverage of MC in STIN is provided by SS-SBS, and we benchmark SC cases, i.e., SC by only a satellite, and SC by only a BS in the TN.}
\label{fig:casestudy}
\end{center}
\vspace{-8mm}
\end{figure}

In this section, we use coverage enhancement as a typical use case and provide simulation results to illustrate the effectiveness of MC in STINs. 
We explore MC for downlink transmission in the SS-SBS architecture and compare it with its single-connectivity (SC) counterparts of single-satellite and single-BS. 
In this setup, a $50$ km$\times 50$ km rural or remote area is considered, where the BSs are distributed with a density of $\lambda_T=6\times 10^{-3}$/km$^2$. The service radius and transmit power of a BS are $R_T=8$ km and $P_T=46$ dBm, respectively. The pathloss exponent in the TN is $\alpha_T=3.5$, and we assume that there are four UEs in each BS's service area, with one of them being the typical UE for simulation analysis.
In the NTN, the satellites are at altitude $H_N=500$ km and are distributed with a density of $5\times 10^{-7}$/km$^2$. The transmit power, main-lobe gain, side-lobe gain, and pathloss exponent are $P_N=50$ dBm, $G_\text{ml}=30$ dBi, $G_\text{sl}=10$ dBi and $\alpha_N\approx 2.0$, respectively. The noise power on the UE side is $-110$ dBm. Both the BS and the satellite are assumed to be equipped with four antennas.

It can be observed from Fig. \ref{fig:casestudy} that the coverage probability of all connectivity cases decreases as the SINR requirement increases. This finding aligns with our expectation, since larger SINR thresholds correspond to stricter requirements for signal quality, and this leads to a reduction in the probability of a successful connection.
From Fig. \ref{fig:casestudy}, it is also shown that the coverage probability of STIN outperforms that of the satellite-only case and that of the TN-only case, particularly when the SINR requirement is low.
This is because MC leverages the complementary strengths of both a satellite in the NTN and a BS in the TN and obtains a combined coverage.
Moreover, as the SINR requirement becomes relatively higher, the coverage probability of STIN remains at least as high as the greater of that of the satellite-only case and that of the TN-only case.
This shows that MC can make full use of combined coverage and try its best to maintain more robust services for the system even under stricter SINR requirements.

\section{Future Research Directions of MC in STINs}
\subsection{Interference Management in MC-Enabled STINs}

\textcolor{black}{
Interference management appears to be a critical challenge in MC-enabled STINs, as more satellite links and/or terrestrial links are added not only to enhance desired signals but also to increase interference signals. A fundamental research question is: how to balance a trade-off between such desired signals enhancement and interference suppression, especially in dense, multi-beam, multi-UE scenarios?
}

\textcolor{black}{
To answer this question, some technical difficulties may arise, such as a joint optimization of adaptive and dynamic coordination of beamforming, power allocation, and user association. Moreover, advanced tools such as ML-based interference prediction and game-theoretic resource competition models can be established to enable dynamic and autonomous interference mitigation.
}

\textcolor{black}{
It can be expected that with these optimization strategies and advanced tools, intelligent interference-aware MC frameworks can be implemented in STINs to improve network capacity and reliability. 
}

\subsection{Massive Access Technique in MC-Enabled STINs}

\textcolor{black}{
Massive access technique enables a satellite or a BS to serve multiple UEs at the same time; while with MC, each UE can connect to multiple satellites and/or BSs. This drastically increases the number of active links, where a key research question arises: how to efficiently manage, allocate, and maintain these increased links while ensuring system scalability and reliability?
}

\textcolor{black}{
Related technical difficulties here include the design of adaptive multiple access mechanisms and advanced successive interference cancellation (SIC) techniques to mitigate interference among coexisting UEs. In addition, load balancing between NTNs and TNs, dynamic clustering of TN UEs and NTN UEs, handover minimization strategies are also vital to prevent congestion and maintain stable connections.
}

\textcolor{black}{
After implementing those strategies, the development of optimized multi-link management frameworks can be expected. These can be helpful for more complex designs of dynamic access protocols, spectral efficiency improvement, and handover disruption minimization.
}

\subsection{Spectrum Sharing Policy between NTN and TN}

\textcolor{black}{
Spectrum sharing policies allow satellites and BSs to access the same spectrum. They are crucial for optimizing bandwidth utilization in MC-enabled STINs, especially when a balance between per-UE performance and overall network capacity needs to be struck. Moreover, MC exacerbates co-channel interference by introducing more wireless links. Furthermore, reserved spectrum allocation should take into account the needs of both NTN and TN equipment, which may vary with UE density and service demands.
}

\textcolor{black}{
To deal with this dilemma, some technical difficulties must be carefully addressed, including the design of dynamic spectrum access mechanisms, monitoring and management of real-time spectrum interference, and fair spectrum allocation between NTN and TN. Advanced techniques, such as reinforcement learning-based policies, can further enhance system adaptability.
}

\textcolor{black}{
With these implementations, the expected results are that dynamic spectrum sharing frameworks that intelligently adjust bandwidth partitioning based on network conditions can be obtained. Other techniques, such as spectrum sensing, can also be integrated to promote efficient and conflict-free coexistence between NTN and TN.
}

\section{Conclusions}
In this article, the architectures, challenges, and typical applications for the MC in STINs are provided. We introduced three general deployment architectures and discuss their characteristics. Then, fundamental challenges in terms of system design, namely satellite networking, beamforming, channel estimation, and synchronization, were explored, each followed by a brief discussion of solutions.
Moreover, we presented four typical applications of MC in STINs and demonstrated their effectiveness through a case study on coverage enhancement.
Finally, we outlined the critical and promising future research directions and hope that this article will provide valuable insights for the real-world design and implementation of MC in STINs towards future generation networks.


\bibliographystyle{IEEEtran}
\bibliography{references.bib}

\begin{thebibliography}{10}
\providecommand{\url}[1]{#1}
\csname url@samestyle\endcsname
\providecommand{\newblock}{\relax}
\providecommand{\bibinfo}[2]{#2}
\providecommand{\BIBentrySTDinterwordspacing}{\spaceskip=0pt\relax}
\providecommand{\BIBentryALTinterwordstretchfactor}{4}
\providecommand{\BIBentryALTinterwordspacing}{\spaceskip=\fontdimen2\font plus
\BIBentryALTinterwordstretchfactor\fontdimen3\font minus \fontdimen4\font\relax}
\providecommand{\BIBforeignlanguage}[2]{{%
\expandafter\ifx\csname l@#1\endcsname\relax
\typeout{** WARNING: IEEEtran.bst: No hyphenation pattern has been}%
\typeout{** loaded for the language `#1'. Using the pattern for}%
\typeout{** the default language instead.}%
\else
\language=\csname l@#1\endcsname
\fi
#2}}
\providecommand{\BIBdecl}{\relax}
\BIBdecl

\bibitem{3gpp.38.821}
{3GPP}, ``{Solutions for NR to support non-terrestrial networks (NTN) (Release 16)},'' {3rd Generation Partnership Project (3GPP)}, Technical Report (TR) 38.821, 05 2021, version 16.1.0.

\bibitem{sun2022integrated}
Y.~Sun, M.~Peng, S.~Zhang, G.~Lin, and P.~Zhang, ``Integrated satellite-terrestrial networks: Architectures, key techniques, and experimental progress,'' \emph{IEEE Netw.}, vol.~36, no.~6, pp. 191--198, Jul. 2022.

\bibitem{pupiales2021multi}
C.~Pupiales, D.~Laselva, Q.~De~Coninck, A.~Jain, and I.~Demirkol, ``Multi-connectivity in mobile networks: Challenges and benefits,'' \emph{IEEE Commun. Mag.}, vol.~59, no.~11, pp. 116--122, Dec. 2021.

\bibitem{majamaa2024toward}
M.~Majamaa, ``Toward multi-connectivity in beyond 5g non-terrestrial networks: Challenges and possible solutions,'' \emph{IEEE Commun. Mag.}, vol.~62, no.~11, pp. 144--150, Jan. 2024.

\bibitem{morosi2024terrestrial}
S.~Morosi, A.~Rago, G.~Piro, F.~Matera, A.~Guidotti, M.~De~Sanctis, A.~V. Coralli, E.~Cianca, G.~Araniti, and L.~A. Grieco, ``Terrestrial/non-terrestrial integrated networks for beyond 5g communications,'' in \emph{Space Data Management}.\hskip 1em plus 0.5em minus 0.4em\relax Springer, Mar. 2024, pp. 89--101.

\bibitem{shang2023coverage}
B.~Shang, X.~Li, C.~Li, and Z.~Li, ``Coverage in cooperative leo satellite networks,'' \emph{J. Commun. Inf. Netw.}, vol.~8, no.~4, pp. 329--340, Dec. 2023.

\bibitem{shang2024multi}
B.~Shang, X.~Li, Z.~Li, J.~Ma, X.~Chu, and P.~Fan, ``Multi-connectivity between terrestrial and non-terrestrial mimo systems,'' \emph{IEEE Open J. Commun. Soc.}, vol.~5, pp. 3245--3262, May 2024.

\bibitem{zheng2024mobile}
J.~Zheng, J.~Zhang, H.~Du, D.~Niyato, B.~Ai, M.~Debbah, and K.~B. Letaief, ``Mobile cell-free massive mimo: Challenges, solutions, and future directions,'' \emph{IEEE Wireless Commun.}, vol.~31, no.~3, pp. 140--147, Feb. 2024.

\bibitem{li2025downlink}
X.~Li and B.~Shang, ``Downlink performance of cell-free massive mimo for leo satellite mega-constellation,'' \emph{arXiv:2501.05655. [Online]. Available: https://arxiv.org/abs/2501.05655}, 2024.

\bibitem{okati2023stochastic}
N.~Okati and T.~Riihonen, ``Stochastic coverage analysis for multi-altitude leo satellite networks,'' \emph{IEEE Commun. Lett.}, vol.~27, no.~12, pp. 3305--3309, Oct. 2023.

\bibitem{omid2024tackling}
Y.~Omid, S.~Lambotharan, and M.~Derakhshani, ``Tackling delayed csi in a distributed multi-satellite mimo communication system,'' in \emph{Proc. Int. Symp. Wireless Commun. Syst. (ISWCS)}, Aug. 2024, pp. 1--6.

\bibitem{li2023channel}
K.-X. Li, X.~Gao, and X.-G. Xia, ``Channel estimation for leo satellite massive mimo ofdm communications,'' \emph{IEEE Trans. Wireless Commun.}, vol.~22, no.~11, pp. 7537--7550, Mar. 2023.

\bibitem{yang2023distributed}
S.~Yang, D.~Wang, L.~Liu, B.~Wang, and C.~Sun, ``Distributed multiple leo satellites cooperative downlink power enhancement transmission scheme based on otfs,'' in \emph{Proc. Int. Conf. Commun. Technol. (ICCT)}.\hskip 1em plus 0.5em minus 0.4em\relax IEEE, Feb. 2024, pp. 1208--1213.

\bibitem{bai2018Multi}
L.~Bai, L.~Zhu, X.~Zhang, W.~Zhang, and Q.~Yu, ``Multi-satellite relay transmission in 5g: Concepts, techniques, and challenges,'' \emph{IEEE Netw.}, vol.~32, no.~5, pp. 38--44, Sep. 2018.

\bibitem{li2024bistatic}
X.~Li, B.~Shang, and Q.~Wu, ``A bistatic sensing system in space-air-ground integrated networks,'' in \emph{Proc. IEEE/CIC Int. Conf. Commun. China (ICCC)}, Aug. 2024, pp. 1823--1827.

\end{thebibliography}

\section*{Biographies}

\begin{IEEEbiographynophoto}{Xiangyu Li}
(xyli@eitech.edu.cn) 
received an M.S. degree in Electrical and Computer Engineering from Georgia Institute of Technology, Atlanta, USA, in 2023, and he is currently pursuing a Ph.D. degree at Shanghai Jiao Tong University (SJTU), Shanghai, China, in the Eastern Institute of Technology (EIT)-SJTU Joint Ph.D. Program. His research interests include space-air-ground integrated networks, non-terrestrial networks, satellite communications, and performance analysis of wireless systems.
\end{IEEEbiographynophoto}

\begin{IEEEbiographynophoto}{Bodong Shang}
(bdshang@eitech.edu.cn) 
received his Ph.D. degree from the Department of Electrical and Computer Engineering at Virginia Tech, Blacksburg, USA, in 2021, and he was a Postdoctoral Research Associate at Carnegie Mellon University, Pittsburgh, USA. 
He is currently an Assistant Professor at the College of Information Science and Technology, Eastern Institute of Technology (EIT), Ningbo, China. 
His research areas are wireless communications and networking, including space-air-ground-sea integrated networks, non-terrestrial networks, and space information networks.
\end{IEEEbiographynophoto}

\end{document}